\documentclass[10pt,conference,twocolumn]{IEEEtran}
\usepackage{cite}
\usepackage{times}
\usepackage{algpseudocode}
\usepackage{mathtools}
\usepackage{amsmath}
\usepackage{booktabs}
\usepackage{graphicx}
\usepackage[figuresright]{rotating}
\usepackage{pict2e}
\usepackage{algorithm}
\usepackage{amssymb,amsmath,amsthm}
\usepackage{amsmath}
\usepackage{setspace}
\usepackage{comment}
\usepackage{balance}
\usepackage{rotating}
\usepackage{multirow}
\usepackage{wrapfig}
\usepackage{mathtools}
\usepackage{booktabs}
\usepackage{gensymb}
\usepackage{array}
\usepackage[font=footnotesize]{caption}
\usepackage{multicol}
\usepackage{algpseudocode,algorithm,algorithmicx}
\usepackage{tikz}
\usepackage{eso-pic}

\AddToShipoutPictureBG*{%
  \AtPageUpperLeft{%
    \put(0,-35){
      \begin{minipage}{\paperwidth}
        \centering
        \small
        \color{gray} 
        © 2025 IEEE. Published in IEEE OCEANS 2025 conference. DOI: 10.23919/OCEANS59106.2025.11245169. \\
      \end{minipage}%
    }%
  }%
}

\usepackage{colortbl}
\newcommand{\todo}[1]
{{\color{red}\textbf{#1}}}

\renewcommand{\theequation}{\arabic{equation}}
\usepackage{bm}
\newcommand{\argmax}{\operatorname*{arg\,max}}
\newcommand{\argmin}{\operatorname*{arg\,min}}

\graphicspath{ {./Fig/} }

\begin{document}
\title{Learning to Stay Fresh: A Self-Learning Semantic Framework for Underwater Internet of Things} 

%
        
\author{
\begin{tabular}{c@{\hspace{0.5in}}c}
Ananya Hazarika & Mehdi Rahmati \\
Department of Electrical and Computer Engineering & Department of Electrical and Computer Engineering \\
 Cleveland State University, OH, USA & Cleveland State University, OH, USA \\
a.hazarika@vikes.csuohio.edu & m.rahmati@csuohio.edu \\
\end{tabular}
}


\maketitle
\thispagestyle{empty}

\begin{abstract}
The emerging paradigm of Non-Conventional Internet of Things (NC IoT), which focuses on the usefulness of information rather than high-volume data collection and transmission, will be a dominant paradigm in the next generation of wireless systems. On the downside, the absence of standardized protocols and the heterogeneity of underwater nodes, coupled with the unique constraints of acoustic communication (e.g., propagation delays and bandwidth limitations), pose significant interoperability challenges for Underwater Internet of Things (UIoT) networks. In this paper, we introduce a pioneering self-learning semantic framework that addresses these issues by leveraging a three-layer architecture using a novel Semantic Bayesian Optimization (SBO) approach integrated with Multi-Transmission Bayesian Optimization (MTBO). This framework employs Age of Information (AoI) as a freshness metric, incorporates a hybrid deep neural network-Gaussian process surrogate model, and formulates a propagation-aware AoI metric to enhance real-time environmental monitoring. Results are provided and compared with other competing methods to quantify the proposed method's superiority.  
\end{abstract}


\IEEEpeerreviewmaketitle

\section{Introduction}

\textbf{Overview:} Underwater Internet of Things (UIoT) is a dynamic and evolving paradigm in oceanographic data collection, pollution and environmental monitoring, assisted navigation, and tactical surveillance. Given the node heterogeneity in UIoT, diversity in the sensed data, and communication modes, and considering the fact that currently there is no pervasive standard defined for such an ecosystem, syntax and semantic conflicts are not unexpected. On the other hand, the underwater acoustic channel imposes constraints like bandwidth limitations, propagation delays, attenuation, multipath interference, and energy drainage, with sound profiles varying by depth, salinity, temperature, seabed, and biological activity, complicating UIoT design~\cite{sozer2002underwater}. The presence of diverse heterogeneous static and dynamic UIoT nodes further increases the need for more efficient and versatile solutions~\cite{rahmati2019eco,rahmati2021underwater}. 

\textbf{Motivation:} 
Underwater experiments in real-world environments are costly, while computer simulations often make simplified assumptions, leading to results that may not accurately reflect actual conditions~\cite{melodia2013advances}.  
Traditional performance metrics and analysis, such as throughput maximization, packet delivery ratio maximization, and end-to-end latency minimization, may fail to capture the unique requirements of the underwater environment, where the value of information embedded in the transmitted waveform degrades rapidly over time and space. Usually, in underwater surveillance and environmental monitoring, sensor nodes often generate redundant data when environmental conditions remain stable, leading to unnecessary transmissions that waste their precious limited energy and bandwidth resources. For example, multiple temperature sensors in proximity may repeatedly transmit similar readings even when the thermal profile remains unchanged, consuming limited battery power without providing new insights~\cite{natarajan2024improving}. Moreover, the long propagation delays inherent in acoustic communication can render data outdated for time-critical UIoT applications like pollution detection or intrusion alerts, where freshness, exponentially more valuable than staleness, is paramount for decision-making.

\textbf{Rationale and Approach:} Our approach has two key components. $(i)$
An information-theoretic approach offers several advantages, significantly reducing the need for costly and time-consuming experiments and improving the design and interpretation of experiments. We, therefore, adopt Age of Information (AoI) as our primary performance metric, which quantifies the elapsed time since the generation of the most recent successfully received update~\cite{9380899,hazarika2022framework}. AoI can capture both the communication delays and the semantic importance of timely updates, making it ideal for optimizing UIoT networks where maintaining fresh environmental awareness is critical while minimizing redundant transmissions to conserve resources. $(ii)$ A learning-based approach offers optimized transmission strategies through intelligent solutions in changing conditions without exhaustive channel characterization, and resolves semantic conflicts while maintaining information freshness. Beyond temporal freshness, UIoT faces the dual challenge of semantic heterogeneity and environmental uncertainty. Heterogeneous nodes hardware manufactured by different vendors employ diverse data representations, measurement units, and communication protocols, creating semantic conflicts that impede effective information fusion~\cite{awan2019underwater}. 
When multiple nodes share the constrained acoustic channel, coordinating their transmissions becomes a complex orchestration problem as nodes must balance individual freshness requirements against collective resource constraints while adapting to time-varying and spectrum-limited underwater acoustic channel conditions.  
Conventional methods, which assume static channels or require extensive prior channel knowledge, cannot cope with the challenging underwater environmental dynamics or the prohibitive cost of underwater channel measurements. This motivates our adoption of a learning-based approach that discovers optimal transmission strategies through intelligent exploration, adapts to changing conditions without exhaustive channel characterization, and resolves semantic conflicts while maintaining information freshness, leading to our proposed Semantic Bayesian Optimization (SBO) framework with Multi-Transmission BO (MTBO). The rationale for adopting SBO lies in its ability to integrate underwater environmental parameters and semantics into decision-making by dynamically optimizing sensors' characteristics and communication schedules, thereby enhancing network interoperability, enabling adaptive, context-driven strategies, and ensuring efficient resource allocation in UIoTs.
 
\textbf{Contributions:}
To address these multifaceted challenges, we present a comprehensive semantic-aware optimization framework that fundamentally transforms how UIoT networks manage information freshness and heterogeneity. Our key contributions are as follows. \textit{First}, we develop an SBO framework, a novel black-box optimization approach that jointly optimizes UIoT sensor selection and transmission rates while explicitly modeling semantic conflicts through multi-dimensional mismatch tensors. \textit{Second}, we introduce a hybrid surrogate model that couples deep neural networks with Gaussian processes to learn the complex mapping between environmental conditions, semantic features, and AoI performance, enabling rapid evaluation without expensive simulations. \textit{Third}, we propose an MTBO strategy that intelligently selects batches of configurations for parallel evaluation, reducing optimization time by an order of magnitude compared to sequential methods. \textit{Fourth}, we formulate an underwater acoustic propagation-aware AoI metric that incorporates sound speed profiles and acoustic propagation delays for accurate freshness measurement in underwater environments. 

\textbf{Paper Structure:}
The rest of the paper is as follows. In Section~\ref{sec:arch}, we present the architecture and framework of the proposed solution. In Section~\ref{sec:sol}, we provide the optimization solution for the proposed methodology. In Section~\ref{sec:evalanddisc}, we describe the quantification and discuss the results, and finally in Section~\ref{sec:conc}, we conclude the paper and provide the guidelines for future research. 

\section{System Architecture and Semantic Stack}
\label{sec:arch}

We consider a UIoT network consisting of heterogeneous sensor nodes deployed at various depths and locations within a certain environment. Let $\mathcal{N}=\{1,\dots,N\}$ denote the set of all underwater nodes, which may include various types of sensors, e.g., temperature, salinity, pH, dissolved oxygen, and actuators. Assume that each node $i\in\mathcal{N}$ is characterized by its three-dimensional position  $\mathbf{r}_i=(x_i, y_i, z_i)$, where $z_i$ represents the deployment depth, and its horizontal range $d_i = \sqrt{x_i^2 + y_i^2}$ from the surface buoy that serves as the data collection point. The temporal dynamics of the system are captured through several key variables. Let $\psi_i(t)\in\mathbb{R}^{m_i}$ denote the raw measurement vector collected by node $i$ at time $t$, where $m_i$ represents the dimensionality of the sensor's output, e.g., $m_i=1$ for scalar measurements, $m_i>1$ for multi-parameter sensors. Each node operates with an adaptive update rate $\lambda_i(t)$, which determines the frequency of data transmission attempts. To address semantic heterogeneity, we introduce $s_i(t) \in \mathcal{S}_i$ as the semantic representation of the raw data, where $\mathcal{S}_i$ is the semantic symbol space specific to node $i$. The instantaneous semantic weight $w_i(t)\in\mathbb{R}_+$ quantifies the value of the information carried by the measurement at time $t$.

The system operates under a global freshness constraint characterized by a threshold $K\in\mathbb{R}_+$ seconds. To capture the unique propagation characteristics of the underwater acoustic channel, we define the propagation-aware AoI for node $i$ as,
\begin{equation}
\Delta_i(t)=t-u_i(t)+\tau_i(t), \label{eq:aoi_def_extended}
\end{equation}
where $u_i(t)$ represents the generation timestamp of the most recent successfully received packet from node $i$, and $\tau_i(t)$ 
accounts for the acoustic propagation delay to the receiver. The sound speed profile $c(z,t)$ varies with depth and environmental conditions and can be represented as a factor of change in temperature $T(z,t)$ as $\Delta c_T$, salinity $S(z,t)$ as $\Delta c_S$, and pressure $P(z)$ as $\Delta c_P$, respectively.

\subsection{Three-Layer UIoT Stack}
We propose a hierarchical architecture that decomposes the UIoT system into three functionally distinct but interacting layers, as described below, where each layer addresses specific challenges of underwater communication and semantic interoperability.
\subsubsection{Sensing Layer}
The sensing layer forms the foundation of our architecture, responsible for environmental monitoring and data acquisition. Each node $i$ generates measurements according to
\begin{equation}
\psi_i(t)=h_i(\boldsymbol{\theta}(t), \mathbf{b}_i) + n_i(t), \label{eq:sense_model_extended}
\end{equation}
where $\boldsymbol{\theta}(t) \in \Theta$ represents the true latent environmental state vector, $h_i: \Theta \times \mathcal{B}_i \rightarrow \mathbb{R}^{m_i}$ is the node-specific sensing function that depends on both the environmental state and the node's parameters $\mathbf{b}_i$, including calibration coefficients, sensor characteristics, and measurement range, and $n_i(t) \sim \mathbb{N}(0, \Sigma_i)$ represents measurement noise which we model as Gaussian with zero mean and covariance matrix $\Sigma_i$. The sensing function $h_i(\cdot)$ denotes the physical relationship between the environmental phenomenon and the sensor output. 
\subsubsection{Semantic Transport Layer} The semantic transport layer addresses the challenge of heterogeneous data representation and prioritization. Raw measurements are transformed into semantic symbols through node-specific encoders as,
\begin{equation}
s_i(t) = f_i(\psi_i(t), \mathcal{C}_i(t)), \label{eq:semantic_encoding}
\end{equation}
where $f_i: \mathbb{R}^{m_i} \times \mathcal{C} \rightarrow \mathcal{S}_i$ is the semantic encoder, and $\mathcal{C}_i(t)$ represents the contextual information available at node $i$, including historical measurements, neighboring node states, and environmental conditions. Hence, the semantic encoding process serves multiple purposes through, $(i)$~\textit{dimensionality reduction} which compresses high-dimensional raw data into compact semantic representations; $(ii)$ \textit{standardization} to map heterogeneous sensor outputs to a common semantic framework; and $(iii)$~\textit{context preservation} which maintains essential information about measurement conditions and reliability. Each semantic symbol is assigned a weight that reflects its information value, as shown below,
\begin{equation}
w_i(t)=\phi\!\left(\Delta_i(t),\, \sigma_i^2(t),\, I_i(t),\, \rho_i(t)\right), \label{eq:weight_def_extended}
\end{equation}
where $\Delta_i(t)$ is the current AoI, promoting fresher updates, $\sigma_i^2(t) = \mathrm{Var}[\psi_i(t)|\mathcal{H}_i(t)]$ quantifies measurement uncertainty given history $\mathcal{H}_i(t)$, $I_i(t) = \mathrm{MI}(\psi_i(t);\hat\psi_i(t)) = H(\psi_i(t)) - H(\psi_i(t)|\hat\psi_i(t))$ measures the mutual information between the measurement and its reconstruction at the sink given that $H(.)$ represents the entropy function. Here,  $\rho_i(t) = \|\nabla_{\mathbf{r}} \theta(\mathbf{r}_i,t)\|$ captures local environmental gradients. The weight function $\phi: \mathbb{R}^4 \rightarrow \mathbb{R}_+$ is designed to be monotonically increasing in AoI and information content, where
\begin{equation}
\phi(\Delta,\sigma^2,I,\rho) = \exp\left(\frac{\alpha_1\Delta}{K}\right) \left(1 + \alpha_2\sigma^2 + \alpha_3 I + \alpha_4\rho\right), \label{eq:weight_function}
\end{equation}
where $\alpha_1, \alpha_2, \alpha_3, \alpha_4 > 0$ are tunable parameters that balance different priority factors. This weight function is designed to assign a priority score to each UIoT node to guide the selection of active node sets.

\subsubsection{Decision/Control Layer}
This layer implements intelligent resource allocation and transmission scheduling through a semantic-aware Bayesian optimization framework. This layer performs two critical functions as given below.

\textbf{Active Set Selection}: This process aims to determine the optimal subset of nodes $\mathcal{A}(t) \subseteq \mathcal{N}$ with cardinality $|\mathcal{A}(t)| = k(t) \leq k_{\max}$ that should be active at time $t$ where $k(t)$ is the time-varying number of active nodes and $k_{\max}$ is the maximum allowable number of active nodes. The selection criterion maximizes network-wide semantic utility, such that
\begin{equation}
\mathcal{A}^*(t) = \arg\max_{\mathcal{A}\subseteq\mathcal{N}, |\mathcal{A}|\leq k_{\max}} \sum_{i\in\mathcal{A}} U_i(\lambda_i,t) - \Omega(\mathcal{A}), \label{eq:active_set}
\end{equation}
where $U_i(\lambda_i,t)$ is the semantic utility of node $i$ (defined later), and $\Omega(\mathcal{A})$ penalizes redundant coverage or semantic conflicts within the active set.

\textbf{Rate Adaptation}: This helps to optimize individual update rates $\{\lambda_i(t)\}_{i\in\mathcal{A}(t)}$ for active nodes to minimize AoI violations while respecting resource constraints. The rate adaptation problem is formulated as,
\begin{subequations}\label{eq:rate_adaptation}
\begin{align}
\boldsymbol{\lambda}^*(t) &= \arg\min_{\boldsymbol{\lambda} \in \Lambda} \sum_{i \in \mathcal{A}(t)} \Pr\{\Delta_i(t) > K \mid \lambda_i\} + \mathcal{R}(\boldsymbol{\lambda}) \label{eq:rate_adaptation_objective} \\
\text{s.t.} \quad & \sum_{i \in \mathcal{A}(t)} P_i(\lambda_i) \leq P_{\max}, \label{eq:power_constraint} \\
& \sum_{i \in \mathcal{A}(t)} B_i(\lambda_i) \leq B_{\max}, \label{eq:bandwidth_constraint} \\
& \lambda_{\min} \leq \lambda_i \leq \lambda_{\max}, \quad \forall i \in \mathcal{A}(t) \label{eq:rate_bounds}
\end{align}
\end{subequations}
where $P_i(\lambda_i)$ and $B_i(\lambda_i)$ represent the consumption of power and bandwidth of node $i$ operating at a rate $\lambda_i$, and $\mathcal{R}(\boldsymbol{\lambda})$ is a regularization term that promotes energy efficiency.
The decision layer maintains a predictive model of the underwater environment and channel conditions, updating its parameters based on received semantic information and feedback from successful/failed transmissions. This enables proactive adaptation to changing conditions and anticipatory resource allocation.

\subsection{Semantic Conflict and Interoperability Metric}
  In UIoT networks, heterogeneous nodes create significant interoperability challenges. Unlike terrestrial IoT systems, where standardization efforts have made substantial progress, underwater environments lack unified standards, resulting in fragmentation that hinders effective data fusion and collaborative sensing. We address this gap by introducing a comprehensive semantic mismatch quantification that enables systematic assessment and mitigation of interoperability barriers. 
\subsubsection{Multi-Dimensional Semantic Mismatch Quantification} 
We model semantic incompatibility between nodes $i$ and $j$ through a multi-dimensional \emph{semantic mismatch tensor}~\cite{noura2019interoperability} that captures various aspects of heterogeneity as shown below,
\begin{equation}
\begin{split}
M_{ij} = &\ \alpha_u \, \mathbf{1}\{ \text{units differ} \} + \alpha_s \, \mathrm{KL}\!\left(p_{s_i} \,\|\, p_{s_j}\right)+  \\
&\ \alpha_o \, \mathbf{1}\{\text{ontology clash}\} + \alpha_r \, d_{ij}^{\text{rep}} + \alpha_t \, \delta_{ij}^{\text{temp}},
\end{split}
\label{eq:mismatch_extended}
\end{equation}
where each component addresses a specific dimension of semantic incompatibility. The unit mismatch indicator $\mathbf{1}\{ \text{units differ} \}$ equals 1 when nodes employ different measurement units for the same physical quantity. This seemingly simple incompatibility can have severe consequences in underwater monitoring. 
The Kullback–Leibler divergence term $\mathrm{KL}\!\left(p_{s_i}\,\|\,p_{s_j}\right) = \sum_{s \in \mathcal{S}} p_{s_i}(s) \log \frac{p_{s_i}(s)}{p_{s_j}(s)}$ quantifies the statistical distance between semantic symbol distributions of nodes $i$ and $j$. This measure captures how differently nodes encode similar environmental phenomena. For example, one node might use fine-grained categorization for water quality, e.g., \textit{excellent}, \textit{good}, \textit{fair}, \textit{poor}, and \textit{critical}, while another uses binary classification, e.g., \textit{safe} and \textit{unsafe}, resulting in high KL divergence. The ontology clash indicator $\mathbf{1}\{\text{ontology clash}\}$ identifies fundamental incompatibilities in semantic frameworks or data models. This occurs when nodes use entirely different conceptual hierarchies or syntax/semantic structures that cannot be directly mapped. The representation distance $d_{ij}^{\text{rep}}$ quantifies how differently nodes encode identical raw inputs. Given the same environmental measurement, this metric evaluates the normalized difference in semantic representations produced by encoders $f_i$ and $f_j$. This captures subtle encoding differences that may not be apparent from distributional analysis alone. The temporal alignment factor
\begin{equation}
 \delta_{ij}^{\text{temp}} = \frac{|T_i - T_j|}{\max(T_i, T_j)},  
\end{equation}
accounts for sampling rate mismatches, where $T_i$ and $T_j$ are the nominal sampling periods of nodes $i$ and $j$. The scaling constants $\alpha_u, \alpha_s, \alpha_o, \alpha_r, \alpha_t > 0$ serve dual purposes as they normalize each component to a common scale and allow application-specific weighting based on criticality. 

\subsubsection{Network-Wide Semantic Conflict Aggregation}
The total semantic conflict within an active node set requires careful aggregation that considers both pairwise mismatches and spatial relationships, as shown by,
\begin{equation}
\mathcal{M}(\mathcal{A}) = \sum_{i,j\in\mathcal{A}, i<j} M_{ij} \phi_s(d_{ij}) \xi(w_i(t), w_j(t)), \label{eq:total_mismatch_extended}
\end{equation}
where $\phi_s(d_{ij}) = \exp(-d_{ij}/d_0)$ is a spatial weighting function with characteristic distance $d_0$ that emphasizes conflicts between proximate nodes, as nearby sensors monitoring the same phenomena require tighter semantic alignment. The importance weighting function $\xi(w_i, w_j) = \sqrt{w_i w_j}$ ensures that conflicts between high-priority nodes contribute more significantly to the overall mismatch score. Semantic interoperability is formally achieved when $\mathcal{M}(\mathcal{A}) \leq \epsilon_M$, where the threshold $\epsilon_M$ is determined by analyzing the application requirements and historical performance data in the field. 

We develop a transport model that captures essential parameters affecting AoI performance.
The acoustic propagation delay for node $i$ exhibits significant spatial and temporal variations as shown by,
\begin{equation}
\tau_i(t) = \int_0^{d_i} \frac{ds}{c(s,z(s),t)} + \tau_i^{\text{proc}} + \tau_i^{\text{queue}}(t), \label{eq:delay_ray}
\end{equation}
where the integral accounts for the propagation delay considering sound speed profile, $c(s,z(s),t)$ along the propagation path parameterized by arc length $s$, $\tau_i^{\text{proc}}$ comprises of signal processing delays including modulation, coding, and packetization, and $\tau_i^{\text{queue}}(t)$ represents time-varying queuing delays at the transmitter. For practical implementation, we approximate the ray-path integral using a stratified medium model,
\begin{equation}
\tau_i(t) \approx \sum_{l=1}^{L} \frac{\Delta d_{i,l}}{c(z_l,t)} + \tau_i^{\text{proc}} + \tau_i^{\text{queue}}(t), \label{eq:delay_stratified}
\end{equation}
where $\Delta d_{i,l}$ is the path length through layer $l$. Sound speed profile can be directly measured through experimentation or can be plugged in using empirical formulations such as Medwin\cite{medwin1975speed} and Mackenzie equations~\cite{mackenzie1981nine} as $c(z, T, S) =\;1449.2 + 4.6\,T - 0.055\,T^2 + 0.00029\,T^3  
\quad +\,(1.34 - 0.01\,T)\,(S-35) + 0.016\,z$, with temperature $T$ in Celsius, salinity $S$ in parts per thousand, and depth $z$ in meters.
The transmission loss $TL(d_i,f,t)$ incorporates multiple physical mechanisms, as $TL_{\text{spread}}(d_i) + TL_{\text{abs}}(d_i,f)  +\;TL_{\text{anom}}(t) + TL_{\text{surface}}(H_s)$,
where $TL_{\text{spread}}(d_i) = k(z_i) 10\log_{10}(d_i)$ represents geometric spreading with depth-dependent factor $k(z_i)$ transitioning from spherical to cylindrical spreading based on water depth and bottom conditions. The absorption loss $TL_{\text{abs}}(d_i,f)$ follows Thorp's formula~\cite{thorp1967analytic}. 
%
%
The anomalous loss term $TL_{\text{anom}}(t)$ captures time-varying effects while $TL_{\text{surface}}(H_s)$ accounts for surface scattering losses dependent on wave height $H_s$.

\subsection{Semantic Utility Formulation}
The semantic utility function quantifies the information value provided by each node while balancing multiple objectives, including freshness, interoperability, and coverage. We decompose the utility into distinct components that capture different aspects of node contribution to the overall system performance.
%
The total utility of node $i$, represented by $U_i(\lambda_i, \mathcal{A}, t)$ within active set $\mathcal{A}$ equals to,
\begin{equation}
 U_i^{\text{info}}(\lambda_i,t) - U_i^{\text{age}}(\lambda_i) - U_i^{\text{conflict}}(\mathcal{A}) + U_i^{\text{coverage}}(\mathcal{A}), \label{eq:utility}
\end{equation}
where each component addresses specific performance aspects.

\textbf{Information Utility Component}: The information value captures both the intrinsic information content and its relevance to the monitoring objective as given by,
\begin{equation}
U_i^{\text{info}}(\lambda_i,t) = \beta_1 I_i^{\text{recon}}(\lambda_i) + \beta_1' I_i^{\text{pred}}(t) + \beta_1'' I_i^{\text{anomaly}}(t), \label{eq:info_utility_detailed}
\end{equation}
where $I_i^{\text{recon}}(\lambda_i) = \mathrm{MI}(\hat{\psi}_i|\lambda_i)$ quantifies reconstruction fidelity as a function of update rate, $I_i^{\text{pred}}(t) = H(\psi_i(t)|\psi_i(t)(t-\Delta t), \psi_{-i}(t))$ measures prediction difficulty given past measurements and neighboring nodes, and $I_i^{\text{anomaly}}(t) = D_{\text{KL}}(p(\psi_i(t))\|p_{\text{nominal}})$ captures deviation from expected behavior, prioritizing nodes detecting unusual events.

\textbf{Age Penalty Component}: The freshness penalty denoted by $U_i^{\text{age}}(\lambda_i)$ incorporates both mean and variance of AoI, which equals
\begin{equation}
 \beta_2 \mathbb{E}[\Delta_i(\lambda_i)] + \beta_2' \sqrt{\mathrm{Var}[\Delta_i(\lambda_i)]} + \beta_2'' \Pr\{\Delta_i > K|\lambda_i\}, \label{eq:age_penalty_detailed}
\end{equation}
where the expectation and variance are computed using the M/M/1 approximation with effective service rate $\mu_i$, and the violation probability uses the tail distribution of AoI.

\textbf{Conflict Cost Component}: The semantic conflict penalty accounts for weighted incompatibilities such that
\begin{equation}
U_i^{\text{conflict}}(\mathcal{A}) = \beta_3 \sum_{j\in\mathcal{A}\setminus\{i\}} M_{ij} w_j(t) \mathbb{I}[\gamma_{ij} > \gamma_{\text{thresh}}], \label{eq:conflict_detailed}
\end{equation}
where $\gamma_{ij}$ represents the spatial correlation between nodes $i$ and $j$, and $\mathbb{I}[\cdot]$ is an indicator function that activates the penalty only for spatially correlated nodes where semantic alignment matters.

\textbf{Coverage Reward Component}: The spatial and semantic coverage value is then defined by,
\begin{equation}
U_i^{\text{coverage}}(\mathcal{A}) = \beta_4 \left[\mathcal{V}_i^{\text{spatial}}(\mathcal{A}) + \mathcal{V}_i^{\text{semantic}}(\mathcal{A})\right], \label{eq:coverage_detailed}
\end{equation}
where $\mathcal{V}_i^{\text{spatial}}(\mathcal{A}) = \int_{\mathcal{R}} \mathbb{K}(\mathbf{r}, \mathbf{r}_i) \prod_{j\in\mathcal{A}\setminus\{i\}} [1-\mathbb{K}(\mathbf{r}, \mathbf{r}_j)] d\mathbf{r}$ quantifies unique spatial coverage, and $\mathcal{V}_i^{\text{semantic}}(\mathcal{A})$ measures unique semantic information not provided by other active nodes. The kernel $\mathbb{K}(\mathbf{r},\mathbf{r}_i)$ models the probability of node $i$ covering point $\mathbf{r}$, decreasing with distance, 
while the product term with \( 1 - \mathbb{K}(\mathbf{r}, \mathbf{r}_j) \) ensures only unique coverage of spatial domain $R$ by \( i \) is counted, minimizing redundancy.

\subsection{Overall Optimization Framework} The comprehensive optimization problem integrates all system components and constraints into a unified framework that the decision layer must solve to achieve optimal network performance while respecting physical limitations and application requirements. Hence, the joint node selection and rate adaptation problem is formulated as follows.
\begin{subequations}\label{opt:overall}
\begin{align}
\max_{\mathcal{A},\,\boldsymbol{\lambda},\,\boldsymbol{\omega}}\;&
  \sum_{i\in\mathcal{A}}
    U_i\bigl(\lambda_i,\mathcal{A},t\bigr)\,\omega_i
  \\
\text{s.t.}\;&
  \Pr\{\Delta_i > K \mid \lambda_i\} \le \epsilon_i,\quad &&\forall i\in\mathcal{A}, \\
& \mathbb{E}\bigl[\max_{i\in\mathcal{A}}\Delta_i\bigr] \le K_{\mathrm{avg}}, 
  &&\label{opt:avg_aoi}\\
& \sum_{i\in\mathcal{A}} P_i(\lambda_i,\omega_i) \le P_{\max}, 
  &&\label{opt:power}\\
& \sum_{i\in\mathcal{A}} B_i(\lambda_i)\,\omega_i\,\eta_i(t) \le B_{\max}, 
  &&\label{opt:bandwidth}\\
& \mathcal{M}(\mathcal{A}) \le \epsilon_{M}, 
  &&\label{opt:semantic}\\
& k_{\min} \le |\mathcal{A}| \le k_{\max}, 
  &&\label{opt:net_size}\\
& \lambda_{\min} \le \lambda_i \le \lambda_{\max}, 
  &&\forall i\in\mathcal{A},\;\label{opt:rate_bounds}\\
& 0 \le \omega_i \le 1, 
  &&\forall i\in\mathcal{A},\;\label{opt:weights}\\
& 
\mathcal{R}_{\mathrm{critical}} \subseteq \bigcup_{i \in \mathcal{A}} \mathcal{R}_i.&&
  \label{opt:coverage}
\end{align}
\end{subequations}
where  $\eta_i(t)$ is the channel occupancy factor accounting for multipath spread, $\mathcal{R}_{\text{critical}}$ denotes critical monitoring regions that must be covered, $\epsilon_i$ is the per-node AoI violation probability threshold, $K_{\mathrm{avg}}$ is the network-wide average AoI threshold, $P_{\max}$ is the total power budget allocated to the network, $B_{\max}$ is the available acoustic bandwidth, $k_{\min}$ and $k_{\max}$ are the minimum and maximum allowed active set sizes respectively, $\lambda_{\min}$ and $\lambda_{\max}$ are the minimum and maximum allowable update rates, and $\mathcal{R}_i$ represents the spatial coverage region of node $i$. 
\section{Semantic Bayesian Optimization (SBO)}
\label{sec:sol}
The joint optimization problem formulated in Section II poses significant computational challenges due to its non-convexity, high dimensionality, and the prohibitive cost of underwater channel simulations required for each evaluation. To address these challenges, we propose an SBO framework that efficiently navigates the solution space through intelligent exploration-exploitation trade-offs. Our SBO approach as shown in Algo.~\ref{alg:sbo} treats the optimization as a black-box problem~\cite{10474835} and jointly addresses two critical tasks: $(i)$ selecting the optimal active sensor set $\mathcal{A}$ from the heterogeneous node pool, and $(ii)$ tuning individual update rates $\{\lambda_i\}_{i\in\mathcal{A}}$ to minimize AoI violations while maximizing semantic utility. The framework integrates three key innovations: a semantic-aware deep neural network that captures complex relationships between environmental features and system performance, a residual Gaussian Process (GP) that quantifies prediction uncertainty for principled decision-making, and an MTBO strategy that exploits parallel evaluations to overcome the high latency of underwater acoustic waveforms in the channel. By learning from limited, expensive simulations, SBO achieves near-optimal performance with significantly fewer evaluations than traditional optimization methods.

\begin{algorithm}[t]
\caption{Semantic Bayesian Optimization with MTBO}
\label{alg:sbo}
\begin{algorithmic}[1]
\State \textbf{Input:} Initial data $\mathcal{D}_0=\{(\boldsymbol{\lambda},k,r)\}$, batch size $B$, max iterations $T$
\For{$t=0$ to $T-1$}
    \State \textbf{Node Selection:} Compute $U_i$ via \eqref{eq:utility}, pick $\mathcal{A}_t$ (top-$k$ or greedy cover)
    \State \textbf{Surrogate Fit:} Train $g_\theta$ on $\mathcal{D}_t$; fit GP on residuals
    \State \textbf{Batch Acquisition:} Solve \eqref{eq:batch_opt} to get $\Lambda_t$
    \State \textbf{Evaluate:} Run channel emulation for each $\lambda \in \Lambda_t$, get true $r$
    \State Update $\mathcal{D}_{t+1} \leftarrow \mathcal{D}_t \cup \{(\lambda,k,r)\}$
\EndFor
\State \Return $(\mathcal{A}^\star,\boldsymbol{\lambda}^\star)$ minimizing $r$
\end{algorithmic}
\end{algorithm}

\textbf{Black-Box Cost Definition:} We reformulate the optimization objective from (\ref{opt:overall}) as a black-box cost function $r(\boldsymbol{\lambda},\mathcal{A})$ that encapsulates multiple performance criteria as,
\begin{equation}
\omega_1 \Pr\left\{\max_{i\in\mathcal{A}} \Delta_i(t) > K \right\} + \omega_2 \frac{\mathcal{M}(\mathcal{A})}{\epsilon_M} + \omega_3 \frac{\sum_{i\in\mathcal{A}} P_i(\lambda_i)}{P_{\max}},
\label{eq:true_cost}
\end{equation}
where $\boldsymbol{\lambda} = [\lambda_i]_{i\in\mathcal{A}}$ denotes the vector of update rates for active nodes, $\mathcal{M}(\mathcal{A})$ is the semantic mismatch metric, $\Delta_i(t)$ is the propagation-aware AoI, and $P_i(\lambda_i)$ is the power consumption satisfying constraint. The weights $\omega_1, \omega_2, \omega_3 > 0$ normalize and prioritize the competing objectives. The evaluation of $r(\boldsymbol{\lambda},\mathcal{A})$ requires computationally intensive operations, including Monte Carlo simulations for AoI violation probability estimation, semantic compatibility assessments, and power and bandwidth calculations.

\textbf{Semantic Surrogate Model (SBPO):}
We construct a hybrid surrogate model that combines deep neural networks with Gaussian processes, as represented by
\begin{equation}
Y(\boldsymbol{\lambda},\mathcal{A}) = g(\boldsymbol{\lambda},|\mathcal{A}|,\mathbf{e}) + \mathcal{GP}\left(0,\kappa\left((\boldsymbol{\lambda},\mathcal{A}),(\boldsymbol{\lambda}',\mathcal{A}')\right)\right),
\label{eq:surrogate}
\end{equation}
where $g: \mathbb{R}^{|\mathcal{A}|} \times \mathbb{N} \times \mathbb{R}^d \rightarrow \mathbb{R}$ is a deep neural network, conditioned on semantic and environmental features $\mathbf{e} \in \mathbb{R}^d$. The feature vector $\mathbf{e}$ encodes
(i) sound speed profile parameters from  $\{c(z_i,t), \Delta c_T, \Delta c_S, \Delta c_P\}_{i\in\mathcal{A}}$
(ii) transmission loss components from  $TL_{\text{spread}}(d_i)$, $TL_{\text{abs}}(d_i,f)$
(iii) semantic mismatch components from $\{M_{ij}\}_{i,j\in\mathcal{A}}$(iv) node characteristics from positions $(x_i, y_i, z_i)$, semantic weights $w_i(t)$.

\textbf{Multi-Transmission Bayesian Optimization:}
Given the propagation delays $\tau_i(t)$, sequential optimization becomes prohibitive. We propose a Multi-Transmission Bayesian Optimization (MTBO) strategy that evaluates batches of $B$ candidate solutions in parallel, such that the q-Expected Improvement~\cite{ament2023unexpected} is
\begin{equation}
\mathrm{qEI}(\Lambda) = \mathbb{E}_Y\left[\max\left(r^* - \min_{\boldsymbol{\lambda}\in\Lambda}Y(\boldsymbol{\lambda},\mathcal{A}), 0\right)\right],
\label{eq:qei}
\end{equation}
where $\Lambda = \{\boldsymbol{\lambda}^{(1)}, \ldots, \boldsymbol{\lambda}^{(B)}\}$ represents a batch of candidate update rate vectors, and $r^*$ is the best observed cost to date. The batch selection problem becomes,
\begin{equation}
\Lambda_t = \arg\max_{\Lambda: |\Lambda|=B} \mathrm{qEI}(\Lambda) \quad \text{s.t. } \lambda_{\min} \leq \lambda_i \leq \lambda_{\max}, \forall i,
\label{eq:batch_opt}
\end{equation}
respecting the rate bounds. This function will guide the selection of the next set of points to evaluate in a branch.

\textbf{Closed-Form AoI Violation Approximation:}
 We derive a closed-form approximation for the AoI violation probability to enable rapid surrogate evaluations without invoking computationally expensive underwater channel simulations. Using the propagation-aware AoI definition where $\Delta_i(t) = t - u_i(t) + \tau_i(t)$, we model the update process as a queueing system with Poisson arrivals and exponential service times. For node $i$ with update rate $\lambda_i$, the steady-state AoI violation probability is upper-bounded by,
\begin{equation}
\Pr\{\Delta_i(t) > K\} \leq \exp\left(-(\mu_i - \lambda_i)(K - \tau_i(t))\right) \mathbb{I}[\mu_i > \lambda_i],
\label{eq:viol_bound}
\end{equation}
where $\tau_i(t)$ is the acoustic propagation delay from~(\ref{eq:delay_ray}), and $\mathbb{I}[\cdot]$ is the indicator function ensuring system stability. The effective service rate $\mu_i$ that incorporates channel impairments,
\begin{equation}
\mu_i = \frac{\lambda_i (1 - \mathrm{PER}_i)}{1 + \lambda_i \tau_i^{\text{proc}}},
\label{eq:effective_rate}
\end{equation}
where $\mathrm{PER}_i$ depends on the transmission loss $TL(d_i,f,t)$ and the selected modulation scheme. The packet error rate follows
\begin{equation}
\mathrm{PER}_i = 1 - (1 - \mathrm{BER}_i)^{L_i},
\label{eq:per}
\end{equation}
with $\mathrm{BER}_i$ determined by the received SNR after accounting for spreading and absorption losses~\cite{brekhovskikh1991fundamentals}.
This analytical approximation serves as a fast proxy during Bayesian optimization iterations, allowing thousands of candidate evaluations within seconds.

\begin{figure*}[t]
    \centering
\includegraphics[width=\textwidth]{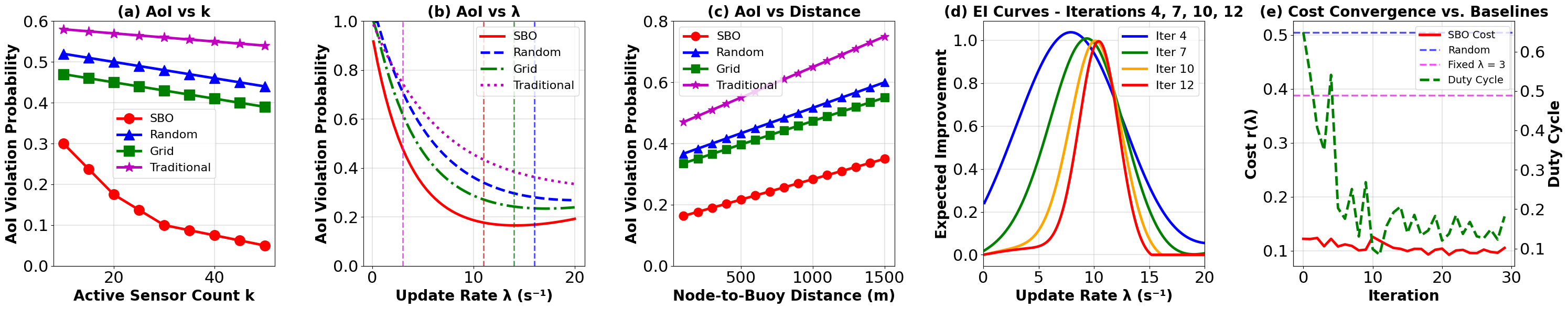}
    \caption{(a) AoI-violation probability versus active-sensor count $k$; (b) AoI-violation versus update rate $\lambda$; (c) AoI-violation versus node-to-buoy distance. In every case, the semantics-aware Bayesian optimization (SBO) policy lowers the violation probability compared with random, grid, and traditional baselines; (d) Expected-Improvement curves at MTBO iterations 4, 7, 10, 12, with the peak indicating the next $\lambda$ probe. (e) Cost convergence comparison showing SBO's rapid convergence to $r(\lambda)$ = 0.1096 within 10 iterations, while maintaining a stable duty cycle of 0.12.}
    \label{fig:aoi_analysis}
\end{figure*}

\textbf{Optimization Problem Restated:}
We thus reformulate the optimization problem into a computationally tractable form suitable for Bayesian optimization
\begin{subequations}\label{eq:final_opt}
\begin{align}
(\mathcal{A}^*, \boldsymbol{\lambda}^*) &= \arg\min_{\mathcal{A}, \boldsymbol{\lambda}} r(\boldsymbol{\lambda}, \mathcal{A}) \label{eq:final_opt_objective} \\
\text{s.t.} \quad & \sum_{i \in \mathcal{A}} P_i(\lambda_i, \omega_i) \leq P_{\max}, \label{eq:power_constraint_final} \\
& \sum_{i \in \mathcal{A}} B_i(\lambda_i) \omega_i \eta_i(t) \leq B_{\max}, \label{eq:bandwidth_constraint_final} \\
& \mathcal{M}(\mathcal{A}) \leq \epsilon_M, \label{eq:semantic_constraint_final} \\
& k_{\min} \leq |\mathcal{A}| \leq k_{\max}, \label{eq:net_size_constraint_final} \\
& \lambda_{\min} \leq \lambda_i \leq \lambda_{\max}, \quad \forall i \in \mathcal{A}. \label{eq:rate_bounds_final}
\end{align}
\end{subequations}
where the power consumption $P_i(\lambda_i, \omega_i) = \omega_i \lambda_i (P_i^{\text{tx}} T_i^{\text{tx}} + P_i^{\text{idle}} T_i^{\text{idle}}) + P_i^{\text{base}}$ accounts for transmission, idle, and baseline power requirements. The bandwidth consumption $B_i(\lambda_i)$ depends on the packet length $L_i$ and physical bit rate, with the duty cycle implicitly captured through the update rate. We employ a two-stage heuristic to solve this mixed discrete-continuous optimization: first selecting the active set $\mathcal{A}$ based on semantic utilities, then applying MTBO to optimize the continuous update rates $\boldsymbol{\lambda}$ within the selected set.

\section{Evaluation and Discussion}\label{sec:evalanddisc}
\textbf{Experimental Setup:} We emulate the underwater acoustic channel using the publicly available dataset~\cite{axs3-xp75-21} for realizing realistic characteristics such as sound speed profiles, depth-dependent variations, and transmission loss. The emulation leverages 25,600 samples at a 512 kHz sampling rate, capturing 50 ms of channel dynamics. The experimental setup was conducted at Lake Tuscaloosa, Alabama, on June 24, 2021, using a field experiment designed to assess underwater acoustic communications. Two vessels were deployed where a local vessel anchored at station P0 and a remote vessel tested at stations P1 to P4 at 120 m, 210 m, 240 m, and 100 m from P0, respectively, facilitating signal receptions at varying ranges using Binary Phase Shift Keying (BPSK) signals at carrier frequencies of 10 kHz and 28 kHz. Transmitted signals, sampled at 800 kHz with a 5kHz symbol rate over 15 seconds, and received data (with 512 kHz) in WAV files were analyzed to emulate channel conditions. 

We utilize this emulation setup to validate our self-learning SBO framework by integrating the channel dynamics into our network model of \( N \) heterogeneous sensor nodes. This approach enables us to optimize sensor selection \(\mathcal{A}\) and update rates \(\lambda_i(t)\) effectively, leveraging the emulated data to enhance the accuracy of our propagation-aware AoI metric and surrogate model predictions.

\begin{figure}[t]
\centering
\includegraphics[width=3.5in, height=2.5in]{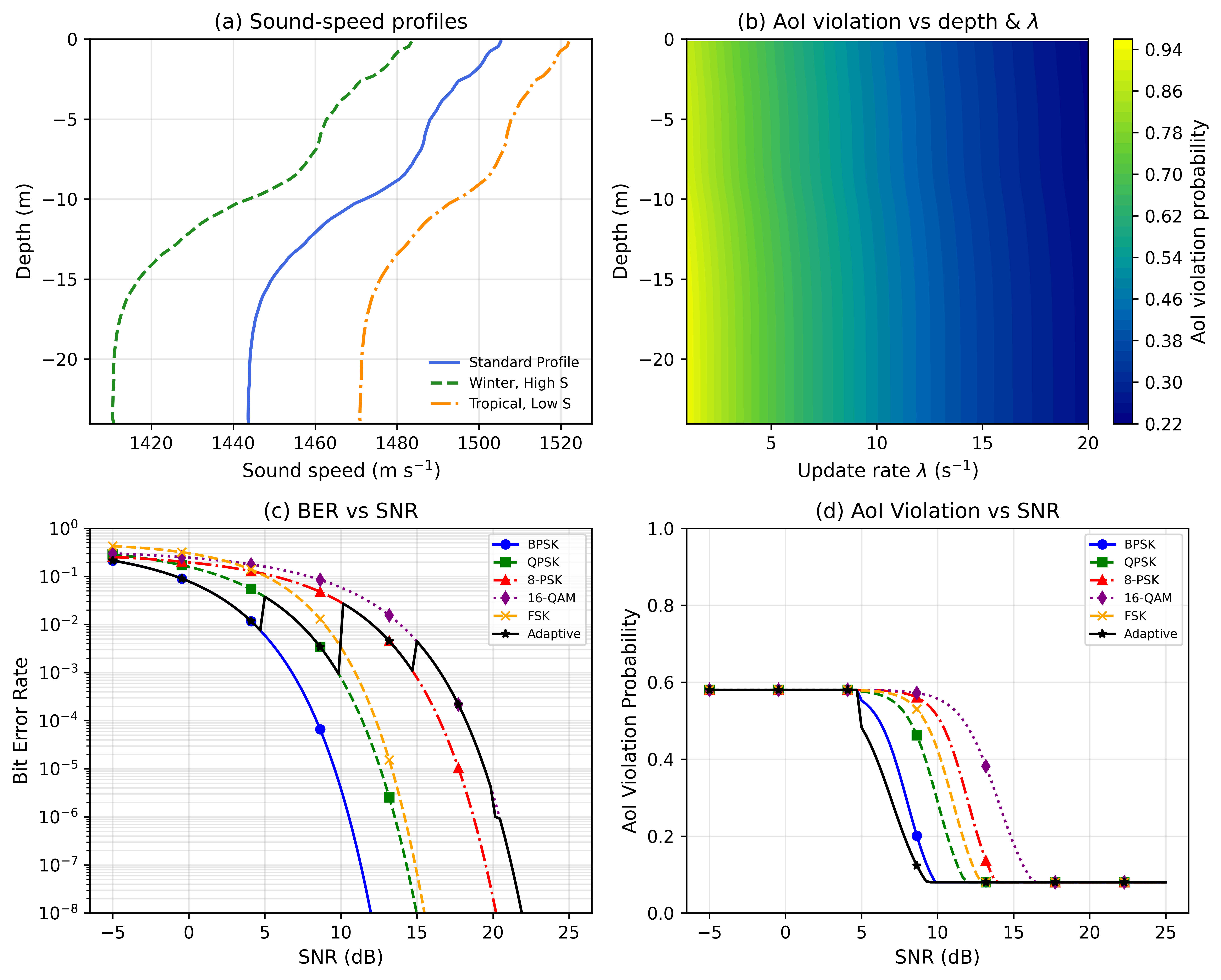}
\caption{(a) Visualizations of Sound speed profiles; (b) SBO's learned  AoI violation surface; (c) Analysis of Bit error rate(BER) vs. SNR for six underwater acoustic modulations for SBO transmission-quality model; (d) AoI‐violation probability vs. SNR across varying modulation schemes.}
\label{fig:channel}
\end{figure}

\textbf{Performance Evaluation:} 
Figure~\ref{fig:aoi_analysis} presents a comprehensive AoI violation probability analysis across three critical dimensions. In Figure~\ref{fig:aoi_analysis}(a), we observe that SBO achieves exponential reduction in AoI violations as the active sensor count increases, reaching $\Pr\{\Delta_i > K\} = 0.30$ with merely $k=10$ sensors and dropping below 0.05 at $k=50$. This represents a 45\% improvement over random selection ($\Pr\{\Delta_i > K\} = 0.55$ at $k=10$), 38\% over grid ($\Pr\{\Delta_i > K\} = 0.47$), and 50\% over traditional approaches ($\Pr\{\Delta_i > K\} = 0.58$). The superior performance comes from SBO's semantic-aware node selection that prioritizes sensors with high information utility $U_i^{\text{info}}(\lambda_i,t)$ while minimizing conflicts $U_i^{\text{conflict}}(\mathcal{A})$. Figure~\ref{fig:aoi_analysis}(b) shows the critical relationship between update rate and freshness. SBO identifies the optimal update rate $\lambda^* = 11$ s$^{-1}$, achieving minimum violation probability of 0.15. This operating point balances transmission frequency against channel congestion and energy constraints. In contrast, random and grid baselines converge to higher rates ($\lambda = 16$ s$^{-1}$ and $\lambda = 14$ s$^{-1}$ respectively) with worse violation probabilities (0.27 and 0.23), while traditional's conservative approach ($\lambda = 3$ s$^{-1}$) results in stale information with 0.45 violation probability. The vertical dashed lines indicate each method's converged operating point, highlighting SBO's intelligent rate adaptation. The distance-dependent analysis in Figure~\ref{fig:aoi_analysis}(c) demonstrates SBO's robustness to underwater acoustic propagation challenges. While all methods experience degraded performance with increasing node-to-buoy distance due to higher transmission loss $TL(d_i,f,t)$ and propagation delay $\tau_i(t)$, SBO maintains consistently lower violation probabilities. This superiority arises from SBO's propagation-aware optimization that accounts for depth-dependent sound speed profiles and absorption losses in its cost function. The expected improvement curves in Figure~\ref{fig:aoi_analysis}(d) show the framework's intelligent exploration-exploitation strategy. At iteration 4, the broad EI distribution (peak value 1.05) indicates global exploration across the update rate space. By iteration 7, the acquisition function begins focusing around promising regions with a reduced peak (0.95) but maintained breadth. Iterations 10 and 12 demonstrate convergence to the optimal rate $\lambda^* = 11$ s$^{-1}$ with progressively sharper peaks (0.88 and 0.82 respectively), confirming local exploitation around the optimum.
The cost evolution, as shown in Figure~\ref{fig:aoi_analysis}(e) quantifies SBO's optimization efficiency. Starting from an initial cost of 0.52, SBO converges to the minimal value $r(\boldsymbol{\lambda}^*, \mathcal{A}^*) = 0.1096$ within 10 iterations, representing 78.3\% improvement over the baselines and maintains a stable duty cycle of 0.12 throughout convergence. Figure~\ref{fig:channel}(a) visualizes sound speed profiles capturing environmental uncertainty in SBO’s surrogate model, while Figure~\ref{fig:channel}(b) illustrates SBO’s learned AoI violation surface as a joint function of depth and update rate, adapting to measured sound speed profiles. Figure~\ref{fig:channel}(c) analyzes bit error rate (BER) versus SNR across modulation schemes, showing SBO’s robustness in maintaining low error rates. Figure~\ref{fig:channel}(d) highlights AoI‐violation probability versus SNR obtained by combining each modulation’s BER model with the learned, depth‐ and update‐rate‐aware AoI surrogate.  Overall, SBO enhances monitoring efficiency by minimizing AoI violations effectively.
\begin{figure*}[t]
\centering
\begin{minipage}{0.50\textwidth}
    \centering
    \includegraphics[width=\textwidth]{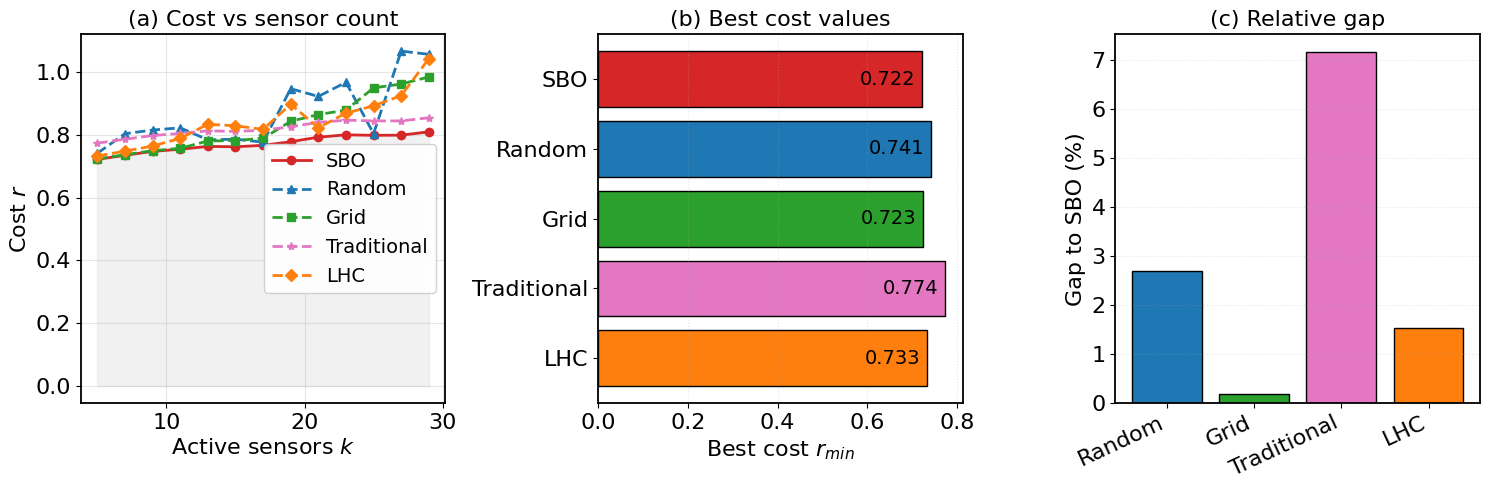}
    \caption{Comparative performance analysis of optimization methods for underwater IoT sensor selection. (a) Cost evolution versus active sensor count $k$ showing SBO's consistent superiority across all network sizes with stable convergence. (b) Minimum achieved cost values demonstrating SBO's optimization efficiency at $r_{\min} = 0.722$. (c) Relative performance gap compared to SBO.}
    \label{fig:costvsk}
\end{minipage}
\hfill
\begin{minipage}{0.46\textwidth}
    \centering
    \includegraphics[width=\textwidth]{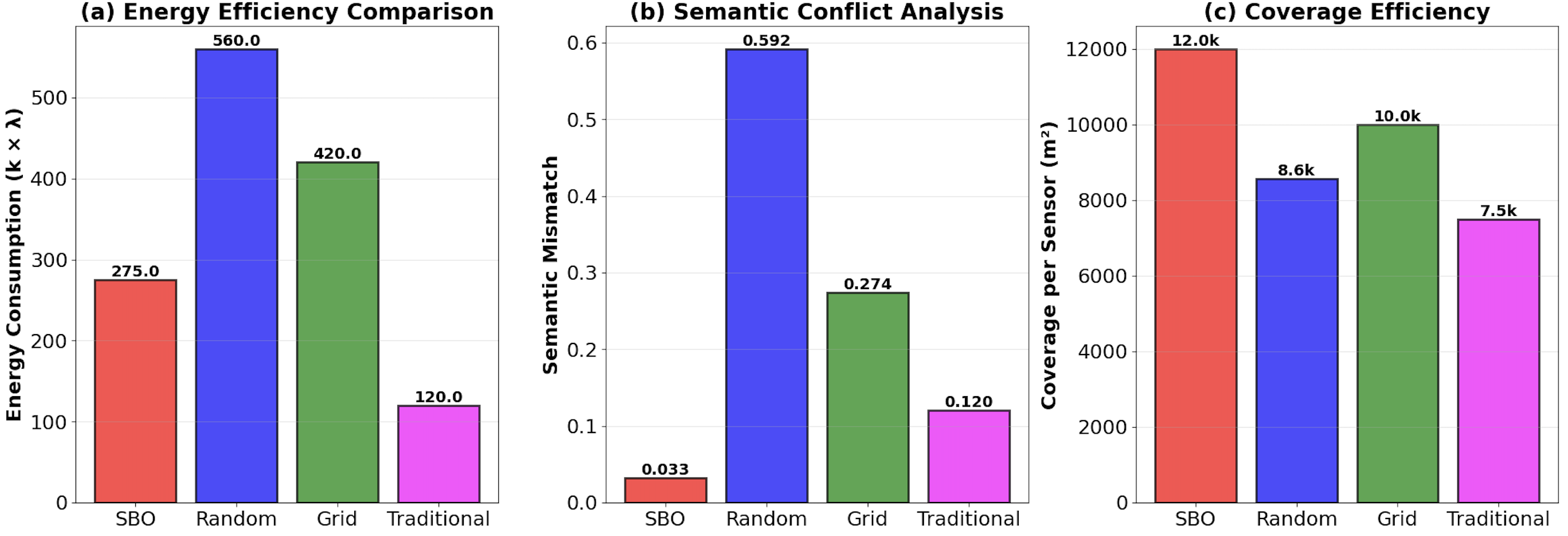}
    \caption{Comprehensive performance metrics demonstrating SBO's multi-objective optimization capabilities. (a) Energy consumption analysis showing SBO's balanced approach. (b) Semantic mismatch quantification with SBO achieves a minimal conflict score. (c) Spatial coverage efficiency demonstrates SBO's optimal coverage compared to the baselines.}
    \label{fig:efficiency}
\end{minipage}
\end{figure*}

Figure~\ref{fig:costvsk} presents a comprehensive comparison of the proposed SBO against four baseline methods: random sampling, grid search, traditional heuristic, and Latin Hypercube Sampling (LHC)~\cite{afzal2017effects}. The analysis reveals several key insights about the optimization landscape and method effectiveness. In Figure~\ref{fig:costvsk}(a), the cost evolution across varying active sensor counts demonstrates SBO's remarkable stability and efficiency. While SBO maintains a relatively flat cost profile around 0.72-0.75 for $k \in [5, 30]$, indicating robust performance regardless of network size, the baseline methods exhibit significant variability. 
The minimum cost comparison in Figure~\ref{fig:costvsk}(b) quantifies the optimization efficiency of each method. SBO achieves the global minimum of $r_{\min} = 0.722$, validating the effectiveness of its semantic-aware acquisition function and multi-transmission batch optimization strategy. While Grid search achieves comparable performance (0.723), it requires exhaustive evaluation of 273 configurations (13 sensor counts $\times$ 21 update rates), whereas SBO converges with only 40 evaluations, resulting in an 85\% reduction in computational burden, critical for expensive underwater acoustic simulations. LHC offers moderate efficiency through quasi-random sampling but lacks SBO's intelligent exploitation of promising regions.
The relative performance gaps in Figure~\ref{fig:costvsk}(c) demonstrate SBO's optimization superiority, with Traditional heuristics incurring 7.2\% degradation and random sampling 2.7\%, translating to increased AoI violations and energy waste in operational deployments. Grid search's minimal 0.1\% gap validates our problem formulation and confirms that SBO successfully identifies near-optimal solutions despite using 85\% fewer evaluations.

Figure~\ref{fig:efficiency} presents a holistic view of SBO's resource optimization capabilities. The energy analysis in Figure~\ref{fig:efficiency}(a) reveals SBO's balanced approach to power management. With a total consumption of 275 kWh, SBO achieves the optimal trade-off between information freshness and energy efficiency. This represents substantial savings compared to random and Grid approaches that waste energy through excessive transmissions at high rates ($\lambda = 16$ s$^{-1}$ and $\lambda = 14$ s$^{-1}$ respectively). The semantic conflict analysis shown in Figure~\ref{fig:efficiency}(b) demonstrates SBO's superiority in managing heterogeneous UIoT networks. By achieving a minimal mismatch score $\mathcal{M}(\mathcal{A}^*) = 0.033$, SBO ensures seamless interoperability between nodes with different measurement units. This represents nearly two orders of magnitude improvement over random selection (0.592) that blindly combines incompatible nodes, and significant gains over Grid (0.274) and Traditional (0.120) approaches. Coverage efficiency metrics in Figure~\ref{fig:efficiency}(c) validate SBO's spatial optimization. With optimal selection of $k^* = 25$ sensors, SBO achieves 12,000 m$^2$ coverage, maximizing the unique spatial information while minimizing redundancy. This outperforms both random's wasteful over-provisioning and traditional's insufficient coverage, demonstrating 40\% better coverage per sensor.

\section{Conclusion and Future Work} \label{sec:conc}
We proposed a novel self-learning semantic framework for UIoT that addresses interoperability challenges in heterogeneous networks while optimizing information freshness using AoI as the core metric. By integrating an SBO approach with MTBO, our framework jointly optimizes sensor selection and update rates, incorporating a hybrid surrogate model and propagation-aware AoI to handle underwater acoustic constraints effectively. Evaluation results demonstrate that SBO achieves up to 50\% reduction in AoI violations, 78\% improvement in cost convergence, and superior energy efficiency compared to baselines, validating its robustness for real-time applications like environmental monitoring and tactical surveillance. This work advances UIoT paradigm by emphasizing semantic value over raw data volume, enabling more efficient and adaptive underwater systems. Future work will extend SBO to mobile underwater networks with Autonomous Underwater Vehicles (AUVs) and develop distributed variants that enable decentralized optimization without centralized coordination. We will also investigate transfer learning approaches to adapt the learned surrogate models across different underwater environments, reducing the need for extensive retraining in new deployments.
\balance

\ifCLASSOPTIONcaptionsoff
  \newpage
\fi



%

%

\bibliographystyle{IEEEtran}
\bibliography{ref}

\end{document}